\documentclass[twocolumn]{webofc}
\usepackage[varg]{txfonts}   
\renewcommand{\vec}[1]{\mbox{\boldmath $#1$}}
\begin{document}
\title{Evolving theoretical descriptions of heavy-ion fusion: 
from phenomenological to microscopic approaches}
%
%

\author{\firstname{} \lastname{K. Hagino}\inst{1,2,3}}

\institute{
Department of Physics, Tohoku University, Sendai 980-8578,  Japan
\and
Research Center for Electron Photon Science, Tohoku
University, 1-2-1 Mikamine, Sendai 982-0826, Japan
\and
National Astronomical Observatory of Japan, 2-21-1 Osawa,
Mitaka, Tokyo 181-8588, Japan}

\abstract{%
We overview the current status of theoretical approaches for heavy-ion 
fusion reactions at subbarrier energies.  
We particularly discuss theoretical challenges 
in the coupled-channels approach, that include i) a description of 
deep subbarrier hindrance of fusion cross sections, 
ii) the role of nuclear dissipation, iii) fusion of unstable nuclei, and iv) 
an interplay between fusion and multi-nucleon transfer processes. 
We also present results of a semi-microscopic approach to heavy-ion 
fusion reactions, that combines the coupled-channels approach with  
state-of-the-art microscopic nuclear structure calculations. 
}
\maketitle
\section{Introduction}
\label{intro}

Fusion is a reaction to form a compound nucleus. 
It plays an important role in several phenomena in nuclear 
physics and nuclear astrophysics, such as synthesis 
of superheavy elements and the energy production in stars. 
In the potential between two colliding nuclei, 
the so called Coulomb barrier is formed by a strong cancellation between 
the Coulomb and a nuclear interactions. 
This Coulomb barrier has to be overcome in order for fusion 
to take place, and thus 
the height of the Coulomb barrier 
defines 
the energy scale of the reaction. In this contribution, 
we mainly consider fusion 
reactions in heavy-ion systems at energies around the Coulomb barrier, 
that is, heavy-ion subbarrier fusion reactions. 

Why are we interested in subbarrier fusion reactions? 
One obvious reason for this is that subbarrier fusion is relevant to 
superheavy elements and nuclear astrophysics. That is, many superheavy 
elements have been synthesized by heavy-ion fusion reactions at 
energies slightly above the Coulomb barrier. Also, it is 
crucially important to understand the dynamics of subbarrier fusion 
reactions in order to make a reliable extrapolation of experimental data 
down to extremely low energies at which fusion reaction takes place in 
stars. 

Besides this obvious reason, there are many other good reasons 
why subbarrier fusion reactions are interesting to study. 
Firstly, 
there is a strong interplay between the nuclear reaction and 
the nuclear structure in there. 
This strong interplay has been well realized 
in a large enhancement of 
fusion cross sections at subbarrier energies 
as compared to 
a prediction of a one-dimensional potential 
model \cite{HT12,DHRS98,BT98,Back14,Canto15}. 
This is in marked contrast to high energy nuclear reactions, in which the
couplings play a much less important role and thus the 
reaction dynamics is much simpler. 

Secondly, 
fusion offers a unique opportunity to study quantum 
tunneling with many degrees of freedom. 
That is, fusion takes place only by quantum tunneling at energies 
below the Coulomb barrier, and the subbarrier enhancement of 
fusion cross sections can be viewed as a result of 
coupling assisted tunneling. 
Heavy-ion fusion reactions are unique in this respect because a variety
of intrinsic degrees of freedom are involved, 
such as static and dynamical 
nuclear deformations with several multipolarities, as well as 
several types of particle
transfer processes with several values of a transfer $Q$-value, 
which can be both 
negative and positive. 
Also, in heavy-ion fusion reactions, 
the incident energy can be easily varied 
in order to study the energy dependence of the tunneling probability, 
whereas the energy is basically fixed in
many other tunneling phenomena in nuclear physics, such as alpha decays.

In order to analyze heavy-ion subbarrier fusion reactions, the 
coupled-channels approach has been developed \cite{HT12,HRK99}, 
which has enjoyed  
a lot of success in reproducing experimental data in many 
reaction systems. In this contribution, we shall first overview the present 
status of this approach and then discuss several remaining theoretical 
challenges. 

\section{Coupled-channels approach}

The field of heavy-ion subbarrier fusion started at 
the late 70's, when a large subbarrier enhancement of fusion cross 
sections was discovered e.g., in the $^{16}$O+$^{154}$Sm system \cite{S78}. 
For this particular system, the enhancement of fusion cross sections 
has been well understood in terms of the deformation of the target nucleus, 
$^{154}$Sm. This nucleus is a typical deformed nucleus with a quadrupole 
deformation parameter of $\beta_2\sim 0.3$. 
When the target nucleus is deformed, the potential between the projectile 
and the target nuclei depends upon the orientation angle, $\theta$, 
of the deformed target. 
The fusion cross sections are then computed as,
\begin{equation}
\sigma_{\rm fus}(E)=\int^1_0d(\cos\theta)\,\sigma_{\rm fus}(E;\theta), 
\label{sigma-def}
\end{equation}
where $\sigma_{\rm fus}(E;\theta)$ is the fusion cross section at the incident 
energy $E$ for a fixed orientation angle $\theta$. 
This formula well reproduces experimental fusion cross sections 
for many systems with a deformed target nucleus, including 
the $^{16}$O+$^{154}$Sm system \cite{HT12,Leigh95}. 

Eq. (\ref{sigma-def}) is valid only when the excitation energy 
of a rotational excitation can be neglected as compared to the curvature 
of the Coulomb barrier \cite{HT12}. In more general cases, one needs to 
solve the coupled-channels equations, 
\begin{equation}
\left[-\frac{\hbar^2}{2\mu}\vec{\nabla}^2+\epsilon_k-E\right]
\psi_k(\vec{r}) +\sum_{k'}\langle\phi_k|V(\vec{r},\xi)|\phi_{k'}\rangle\,\psi_{k'}(\vec{r})=0,
\label{cc}
\end{equation}
where $\mu$ is the reduced mass for the relative motion 
between the colliding nuclei, 
$\phi_k$ is the intrinsic wave function, for which 
$\epsilon_k$ is the excitation energy. 
$V(\vec{r},\xi)$ is the total potential, which includes both the bare 
and the coupling potentials, $\xi$ denoting the intrinsic 
coordinate. Notice that the coupled-channels equations, 
Eq. (\ref{cc}), are derived  
by expanding the total wave function, $\Psi$, with the basis functions 
$\phi_k$, as, 
\begin{equation}
\Psi(\vec{r},\xi)=\sum_k\psi_k(\vec{r})\,\phi_k(\xi). 
\label{expansion}
\end{equation}

Important ingredients for the coupled-channels approach are the internuclear 
potential, $V$, as well as the nature of the intrinsic degrees of 
freedom, $\xi$. For the latter, one often employs the macroscopic collective 
model \cite{HT12}. Such approach has successfully accounted for experimental 
data for many systems, and has been a standard tool in analyzing experimental 
data for heavy-ion subbarrier fusion reactions \cite{HRK99}. 
The coupled-channels approach also offers a natural explanation for the 
barrier distribution \cite{RSS91}, which is intimately related to the 
eigenchannel representation of the coupled-channels equations \cite{HT12}. 

\section{Remaining challenges in the coupled-channels approach}

\subsection{Deep subbarrier hindrance of fusion cross sections}

Despite its success, there are still several theoretical challenges 
in the coupled-channels approach to subbarrier fusion. 
In this section, we discuss four main challenges. 

The first problem which we consider 
is the hindrance phenomena of fusion cross sections 
at deep subbarrier energies \cite{Back14}. 
That is, even when  
a standard coupled-channels calculation well 
reproduces experimental fusion cross sections 
in the vicinity of the Coulomb barrier, it appears that such calculation 
tends to overestimate fusion cross sections at deep subbarrier energies. 
Even though the exact origin for this phenomenon has not yet been clarified, 
so far there are mainly two models which account for the deep subbarrier 
hindrance of fusion cross sections. 
One is based on the sudden approach, in which fusion reaction 
is assumed to take place so fast that the density distribution of each 
colliding nucleus is frozen during fusion \cite{Back14,ME06}. 
The frozen density approximation leads to a repulsive core in 
an internucleus potential, which results in a shallow pocket. 
High angular momenta are cut-off when a potential is shallow, and this 
angular momentum cut-off is the main cause of deep subbarrier hindrance 
in this model. The second model, on the other hand, is based on the 
adiabatic approach, in which reaction is assumed to take place 
so slowly that the density distribution is optimized at every 
instant \cite{IHI09,I15}. 
This results in a deep and thick internucleus potential, and 
a tunneling of such thick potential has a responsibility to hinder fusion 
cross sections. 

So far, both models have been equally successful in reproducing the observed 
fusion hindrance phenomenon. In order to disentangle them, it will be 
necessary to properly model the dynamics, such as the energy dissipation, 
around and after the touching point of the 
colliding nuclei \cite{IHI07}, for which most of the current 
approaches resort either 
to the incoming wave boundary condition \cite{HT12} or to a short 
range imaginary potential. 
Such modeling will be important also to understand the dynamics of 
quasi-fission and fusion reactions relevant to superheavy nuclei. 

\subsection{Fusion above the barrier}

\begin{figure}[tb]
\centering
\includegraphics[clip,width=7.5cm]{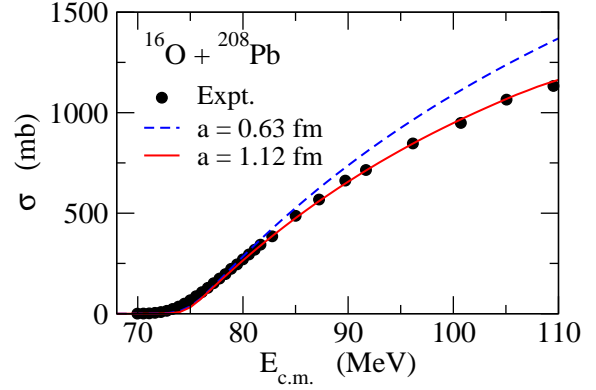}
\caption{
Fusion cross sections for the $^{16}$O+$^{208}$Pb system. 
The dashed and the solid lines denote the results of single-channel 
calculations with a Woods-Saxon internuclear potential with the 
surface diffuseness parameter of $a=0.63$ and 1.12 fm, respectively. 
The experimental data are taken from Ref. \cite{Morton99}. }
\end{figure}

We next discuss fusion cross sections at energies above 
the Coulomb barrier. It has been a long standing problem that 
a Woods-Saxon internuclear potential with 
a standard value of surface diffuseness parameter, that is, $a\sim0.63$ fm, 
systematically overestimates fusion cross sections above the Coulomb 
barrier \cite{Newton04}. If the surface diffuseness parameter is 
phenomenologically increased, experimental data appear to be 
reproduced (see Fig. 1 for a typical example of the situation, that is, the 
fusion cross sections for the $^{16}$O+$^{208}$Pb system), but 
the exact origin of such surface diffuseness anomaly has not yet been 
understood. 

\begin{figure}[tb]
\centering
\includegraphics[clip,width=7.5cm]{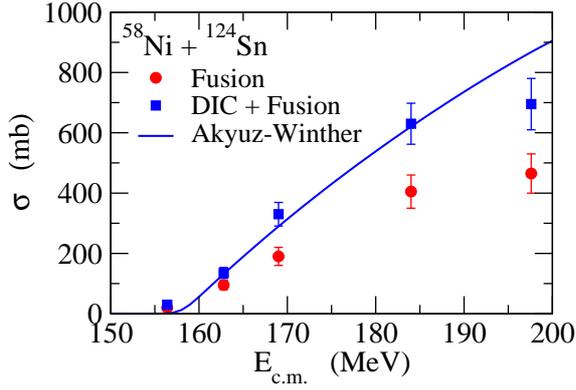}
\caption{The experimental data for 
fusion cross sections (the filled circles) and 
a sum of fusion and deep-inelastic collision cross sections 
(the filled squares) for the $^{58}$Ni+$^{124}$Sn system. These 
data are taken from Ref. \cite{Wolfs87}. 
The solid line shows the result of a single-channel calculation with the 
Aky\"uz-Winter potential \cite{AW}. }
\end{figure}

The situation shown in Fig. 1 appears similar to the one in 
heavier systems where deep-inelastic collision well competes with fusion. 
In Ref. \cite{R94}, Reisdorf argued that 
a barrier passing 
calculation well accounts for a sum of fusion and deep-inelastic 
cross sections e.g., for the $^{58}$Ni+$^{124}$Sn system, even though 
it overestimates fusion cross sections themselves (see Fig. 2). 
The overestimate of fusion cross sections shown in Fig. 1 
may have a similar origin \cite{Newton04}. 

It has been known that energy and angular momentum dissipations play 
an important role in deep inelastic collisions. 
Therefore, in order to resolve the surface diffuseness anomaly, it is 
crucially important to understand the nuclear dissipation in heavy-ion 
reactions. So far, current fusion models are all ``friction free'', in a 
sense that friction is taken into account only as a strong absorption 
well inside 
the barrier. It will be important to extend it by including the effect of 
dissipations at larger distances. 
Such model will provide a quantal model for deep inelastic 
collisions, and at the same time it will also describe dissipative tunneling 
in heavy-ion fusion reactions \cite{TH17}. 

\subsection{Fusion of unstable nuclei}

Fusion of unstable nuclei has been discussed for 
some time \cite{Canto15,Canto06,Hagino00,Diaz-Torres02}, 
but its underlying dynamics has not yet been 
completely understood. For unstable nuclei, the breakup of 
projectile nucleus becomes important because of 
its weakly bound property. This process complicates the whole fusion 
process, especially when one intends to separate between complete 
and incomplete fusion cross sections \cite{Canto15,Canto06}. 
Furthermore, transfer processes may also significantly affect the dynamics 
of fusion reactions of weakly bound nuclei \cite{Raabe04,Lemasson09,Luong13}. 
It is therefore important for any theoretical calculation to take into 
account fusion, breakup, and transfer processes simultaneously 
in order to model fusion of unstable nuclei. 
Such calculation would require a large scale computing, 
and has been rather scarce. An exception is a time-dependent wave 
packet approach \cite{Yabana06}, which however has been limited only to total 
fusion cross sections. Because of this, the enhancement of fusion 
cross sections observed e.g., in the $^{15}$C+$^{232}$Th system \cite{A11}, 
as compared to fusion cross sections for the $^{12,13,14}$C+$^{232}$Th systems, 
has remained theoretically unexplained. 

\subsection{Interplay between fusion and transfer}

The nucleon transfer processes, especially the two-neutron 
transfer process, are important subjects not only in subbarrier fusion 
but also in connection to 
the pairing correlation in neutron-rich 
nuclei \cite{pair01,HS05,PBM11}. 
The multi-neutron transfer process \cite{CPS09} 
would play an important role also in 
fusion of neutron-rich skin nuclei, 
in order to reach the island of stability 
in the superhevy region. 

\begin{figure}[tb]
\centering
\includegraphics[clip,width=7.5cm]{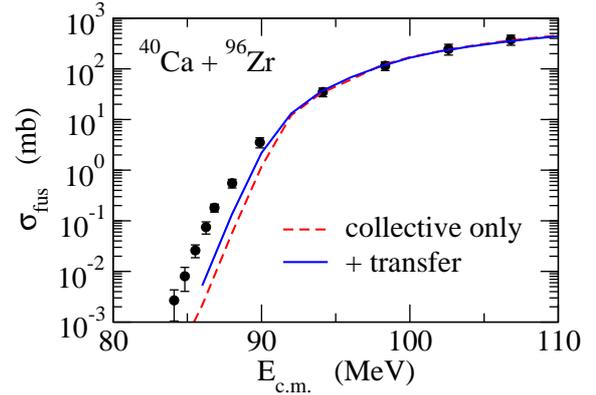}
\caption{Fusion cross sections for the 
$^{40}$Ca+$^{96}$Zr system. The dashed line is obtained by 
including only the collective octupole phonon excitations in the 
projectile and the target nuclei, while the solid line is obtained 
by including in addition the multi-neutron transfer channels. 
The strengths of the transfer couplings are determined so as to reproduce 
the experimental transfer cross sections for this system. 
The experimental data for fusion cross sections 
are taken from Ref. \cite{Stefanini14}.} 
\end{figure}

An important theoretical 
issue here is to reproduce experimental fusion and transfer 
cross sections simultaneously. Recently, we have investigated this problem 
for the $^{40}$Ca+$^{96}$Zr system \cite{SH15}. 
By including the multi-neutron transfer process in the coupled-channels 
approach, we have found that the coupling strengths for the transfer 
couplings, which reproduce the transfer cross sections, largely underestimate 
fusion cross sections (see Fig. 3). To this end, we have included 
one octupole phonon excitation in $^{40}$Ca, the octupole phonon excitations 
in $^{96}$Zr up to the three-phonon levels, and the 
multi-nucleon transfer process up to three-neutron transfer, with 
both simultaneous and direct two-neutron transfer couplings \cite{SH15}. 
We have assumed a transfer to a single effective channel for each 
transfer partition, and set its energy to be the same as the optimum 
$Q$-value, $Q=0$. 
Following Ref. \cite{EL89}, we have assumed that the properties of 
the collective excitations 
do not change even after the transfer. 
A striking fact is that the transfer process enhances fusion 
cross sections only a little if the coupling strengths are chosen so 
as to reproduce the transfer data. This indicates that 
the coupling scheme which has been employed in this calculation needs 
a further 
extension, by taking into account e.g., 
a distribution of transfer $Q$-value as well as 
changes in collective states after 
transfer. 
We have reached a similar conclusion also for 
the $^{40}$Ca+$^{58,64}$Ni systems \cite{SBH16}. 

Very recently, it has been found experimentally that 
the multi-nucleon transfer 
processes in $^{16,18}$O, $^{19}$F + $^{208}$Pb reactions populate highly 
excited states in the target-like nuclei \cite{Rafferty16}. 
A population of high excited states would lead to energy dissipation. 
It would be another challenging problem to describe such process 
within the framework of coupled-channels approach \cite{TH17}. 

\section{Semi-microscopic approach to heavy-ion fusion reactions}

\subsection{Current status of fusion modelings} 

\begin{figure}[tb]
\centering
\includegraphics[clip,width=8.2cm]{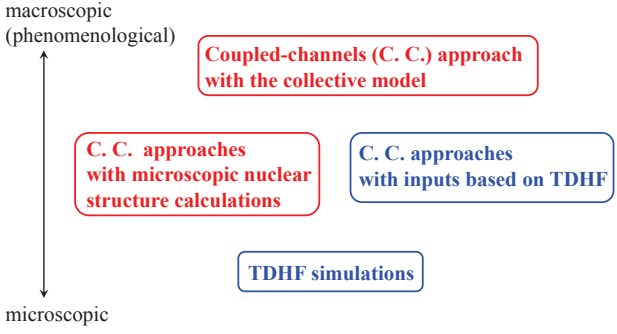}
\caption{The current status of 
theoretical approaches to 
heavy-ion subbarrier fusion reactions. }
\end{figure}

Figure 4 summarizes the current status of 
theoretical approaches to 
heavy-ion subbarrier fusion reactions. 
The coupled-channels approach with the collective model is 
categorized as a macroscopic and phenomenological approach, 
which needs empirical information on the coupling strengths 
and an internuclear potential. 
On the other hand, a full time-dependent Hartree-Fock (TDHF) 
simulation, 
which has rapidly been developed in the past few 
years \cite{Simenel12,Sekizawa13,Scamps12}, 
does not require any empirical input once the energy functional is 
specified. This is a microscopic approach, starting from the 
nucleon degree of freedom. A big challenge in this approach, however, 
is a well known fact that TDHF cannot describe quantum tunneling, 
and thus heavy-ion fusion reactions at energies below the Coulomb 
barrier. One would have to go beyond the mean-field approximation 
in order to resolve this problem, e.g., by using the time-dependent generator 
coordinate method (TDGCM) \cite{Reinhard83}. No realistic calculation, 
however, 
has been done so far 
because such calculation is still quite numerically expensive. 

In between the macroscopic and the full microscopic approaches, one can 
also think about semi-microscopic approaches. These are coupled-channels 
calculations with inputs either from a microscopic nuclear 
structure calculation \cite{HY15,IM13}
or from a TDHF simulation \cite{UO08,WL08,SDHW13}. 
The density-constrained TDHF (DC-TDHF) method \cite{UO08} 
is categorized into this group. 
These semi-microscopic approaches can be applied to fusion at energies 
below the Coulomb barrier, as one solves a quantum mechanical equation 
once it is set up using microscopic inputs. 

\subsection{Coupled-channels calculations with 
microscopic nuclear structure calculations} 

As an example of the semi-microscopic approaches to heavy-ion 
subbarrier fusion reactions, we present in this subsection 
the approach which combines 
the coupled-channels calculations with the 
multi-reference covariant 
density functional theory (MR-CDFT) 
for nuclear collective excitations \cite{HY15}. 
In this approach, the nuclear potential in the 
coupled-channels equations, Eq. (\ref{cc}), 
is assumed to be a deformed Woods-Saxon type with a 
microscopic multipole operator, 
$Q_{\lambda\mu}=\sum_ir_i^\lambda Y_{\lambda\mu}(\hat{\vec{r}}_i)$. 
That is, 
\begin{equation}
V(r,\xi)=\frac{-V_0}{1+\exp\left(
\frac{r-R_0-\sqrt{\frac{2\lambda+1}{4\pi}}\,R_T\alpha_{\lambda 0}}{a}
\right)}
\label{pot}
\end{equation}
with 
\begin{equation}
\alpha_{\lambda\mu}=\frac{4\pi}{3e}\,\frac{1}{Z_TR_T^\lambda}\,Q_{\lambda\mu},
\end{equation}
where $R_T$ and $Z_T$ are the radius and the atomic number 
of the target nucleus, respectively (we here consider 
excitations in the target nucleus). 
In writing Eq. (\ref{pot}), 
we have used the isocentrifugal 
approximation \cite{HT12}. 
The Coulomb coupling is also taken into account in a similar fashion. 
The wave functions for the collective states, $\phi_k(\xi)=\phi_{IM}(\xi)$, 
in Eq. (\ref{expansion}) are many-body wave functions 
generated from microscopic 
nuclear structure calculations, such as the 
multi-reference covariant 
density functional theory. These microscopic wave functions 
also yield all the matrix elements of 
the multipole operator, 
$\langle\varphi_{I0}|Q_{\lambda0}|\varphi_{I'0}\rangle$. 
The coupled-channels equations can then be constructed 
in a similar way as in the 
macroscopic approach for a given internuclear potential. 

In heavy-ion fusion reactions, the coupling between the ground state 
and the lowest-lying 
collective state play the most important role, even though the couplings 
from the lowest-lying state to higher states are often important as well. 
Since the strength for the coupling between the ground state and 
the lowest-lying state can 
often be estimated from an experimental transition probability, 
we introduce an overall scaling factor to all the matrix 
elements so that the transition from the lowest-lying collective 
state to the ground state is consistent with experimental data. 
The MR-CDFT calculation then provides 
the relative strengths among collective levels, which are often not 
available experimentally. 
The excitation energy, on the other hand, is often known well 
for many levels, and we simply use them in the calculations 
whenever they are available. 

\begin{figure}[tb]
\centering
\includegraphics[clip,width=7.5cm]{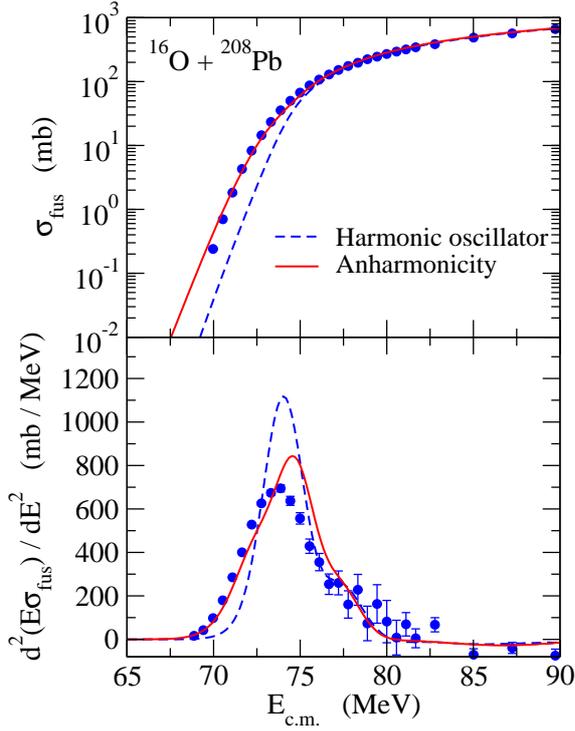}
\caption{
The fusion cross sections (upper panel) and the fusion
  barrier distribution (lower panel) for the $^{16}$O+$^{208}$Pb
  system obtained with the semi-microscopic coupled-channels 
calculation (the solid line). 
For this purpose, the multi-reference covariant 
density functional theory is employed, and states 
up to ``two-phonon'' configurations 
are included in the calculation. 
The dashed line shows the results of the two-phonon coupling in
the harmonic oscillator limit. 
The experimental data are taken from Ref.~\cite{Morton99}.}
\end{figure}

Figures 5 shows the fusion 
cross section $\sigma_{\rm fus}(E)$ 
and the fusion barrier distribution 
$D_{\rm fus}(E) = d^2(E\sigma_{\rm fus})/dE^2$ \cite{DHRS98,RSS91}
for the $^{16}$O+$^{208}$Pb reaction so obtained. 
The dashed line shows the result of the 
coupled-channels calculations including up to the double quadrupole 
and octupole 
phonon states (that is, 
3$_1^-$, 2$_1^+$,
$(3_1^-)^2$, $(2_1^+)^2$, and $3_1^-\otimes 2_1^+$ states) 
in $^{208}$Pb 
in the harmonic oscillator limit. 
As shown in the figure, 
this calculation 
overestimates the height of the main peak in the barrier
distribution, as in many previous calculations 
\cite{Morton99,Esbensen07,YHR12}.
The result of the semi-microscopic coupled-channels calculations 
is shown by the solid line. For this purpose, we generate the collective 
states with the MR-CDFT approach with the PC-PK1 interaction \cite{PC-PK1}. 
In the coupled-channels calculation, in addition to the entrance channel,
we include the one-octupole phonon state, 3$^-_1$, at 2.615 MeV,
the ``one-quadrupole'' phonon state, 2$^+_1$, and several states which are
strongly coupled to those 3$^-_1$ and 2$^+_1$ states by the octupole and
the quadrupole couplings. 
The whole two-octupole-phonon candidate
states are included in this model space.
It is remarkable that the semi-microscopic calculation yields 
a much lower
main peak in the fusion barrier distribution, 
and the agreement with the experimental data 
is considerably improved 
both for the fusion
cross sections and for the barrier distribution.

For this good reproduction, we find that 
the coupling between the 3$^-_1$ and
the 2$^+_1$ states play an important role. 
The couplings between the two-octupole-phonon 
states and the excited negative parity states also play a role. 
In the calculations in the harmonic oscillator limit, 
the 3$^-$, 2$^+_1$ and the 5$^-_1$ states are treated as
independent phonon states, 
and the couplings among those states are absent. 
In contrast,
in the present semi-microscopic calculation, 
the 2$_1^+$ state 
has in part the two octupole phonon character, $(3^-)^2$. 
Likewise, the 1$^-$ and $5_1^-$ states have both
the $(3^-)^3$ and the $3^-\otimes 2^+$ characters. 
Apparently those anharmonicity
effects in the transition strengths
lead to the strong couplings between the ground state and those
states via multiple octupole excitations, significantly improving the
previous coupled-channels calculations.

\section{Summary} 

Heavy-ion subbarrier fusion reactions show a strong interplay 
between nuclear 
reaction and nuclear structure, and contain a variety of rich 
physics, such as coupling assisted tunneling and energy dissipation. 
The coupled-channels approach with a macroscopic and phenomenological 
description for 
nuclear structure 
has been developed in order to understand the 
dynamics of subbarier fusion. 
We have argued in this paper that 
a theoretical description of 
subbarrier fusion is now gradually being shifted from the phenomenological 
approach to more microscopic modelings. A full time-dependent Hartree-Fock 
simulation is a good example for this, 
even though this approach still has a difficulty 
in applying to fusion at energies below the Coulomb barrier. 
We have also developed the semi-microscopic coupled-channels approach, 
which uses inputs generated from microscopic nuclear structure calculations. 
We have shown the result of such calculation for the $^{16}$O+$^{208}$Pb 
system, which has considerably improved the previous coupled-channels 
calculations in the harmonic oscillator limit. 
This approach is more flexible than the conventional 
approach for the coupled-channels calculations, because it can be applied 
also to transitional nuclei, which show neither the vibrational nor the 
rotational characters. 

There are still many theoretical challenges in heavy-ion 
subbarrier fusion reactions, for which 
dissipation and multi-nucleon transfer 
make two important keywords. These are: i) deep subbarrier hindrance 
of fusion cross sections, ii) fusion above the barrier and 
the role of 
nuclear dissipation, iii) fusion of unstable nuclei, and iv) an interplay 
between fusion and transfer. 
Microscopic and semi-microscopic approaches 
to subbarrier fusion will hopefully resolve some of these challenges in a near 
future. 

\bigskip

We thank J.M. Yao and G. Scamps for collaborations 
and for useful discussions.

\end{document}